\documentstyle[preprint,prc,aps,psfig]{revtex}

\renewcommand{\vec}[1]{\mbox{\boldmath $#1$}}

\tightenlines

\begin{document}

\title{Role of virtual break-up of projectile \\ 
in astrophysical fusion reactions}
\author{K. Hagino\footnote{On leave from: 
Yukawa Institute for Theoretical Physics, Kyoto
University, Kyoto 606-8502, Japan }, 
M.S. Hussein 
\footnote{
Permanent address: Instituto de Fisica, 
Universidade de Sao Paulo, CP 66318, 05315-970, Sao Paulo, SP, Brazil},
and A.B. Balantekin
\footnote{Permanent address: Department of Physics, University of Wisconsin, 
Madison, Wisconsin 53706}
}
\address{Institut de Physique Nucl\'eaire, IN2P3-CNRS, \\ 
Universit\'e Paris-Sud, F-91406 Orsay Cedex, France}

\maketitle

\begin{abstract}

We study the effect of virtual Coulomb break-up, commonly known as the dipole 
polarizability, of the deuteron 
projectile on the astrophysical fusion reaction $^3$He($d,p$)$^4$He. 
We use the adiabatic approximation to estimate the potential shift 
due to the $E1$ transition to the continuum states in the deuteron,  
and compute the barrier penetrability in the WKB approximation. 
We find that the enhancement of the penetrability due to the 
deuteron break-up is too small to resolve the longstanding puzzle 
observed in laboratory measurements 
that 
the electron screening effect 
is surprisingly larger than 
theoretical prediction based on an atomic physics model. 
The effect of the $^3$He break-up in the $^3$He($d,p$)$^4$He reaction, as 
well as the $^7$Li break-up in the $^7$Li($p,\alpha$)$^4$He reaction  
is also discussed. 

\end{abstract}


The problem of electron screening effect on nuclear fusion reactions 
measured at a laboratory at very low incident energies has not yet 
been fully understood. 
The rise of the astrophysical $S$-factor for reactions 
such as $^3$He($d,p$)$^4$He and $D(^3$He, $p$)$^4$He as the incident 
energy goes down below about 50 keV has been attributed to the screening 
effect of the bound electrons in the target atom (or molecule), which 
shields the Coulomb potential between the colliding 
nuclei \cite{ALR87,EKN88}. 
It has been found, however, that the amount of the enhancement 
of the $S$-factor 
can be accounted for only when an unrealistically large 
electron screening energy $U_e$ is used in a calculation
\cite{SRSC01,L93,A98,ARG01,CFJ00}. 
The value of $U_e$ required to fit the data 
ranges between 0.88 and 14.5 times the adiabatic value (see 
Table 1 in Ref. \cite{FRVR03} for a summary), where 
the screening energy is given by a difference of electron binding energies 
between the unified and isolated systems \cite{ALR87,B90}. 
Since the adiabatic approximation should provide the upper limit of the 
screening energy \cite{THAB94,THA95,BT98,SKLS93}, the mechanism of the low 
energy enhancement in the $S$-factor remains an open problem. 
A noteworthy recent paper is by Barker, who refitted the experimental 
data by including the screening 
correction as a free parameter and obtained smaller screening 
energies which are consistent with the adiabatic value in many 
systems \cite{B02}. However, for some cases, the optimum screening 
energy still exceeds the adiabatic value, and the fitted $S$ factors 
are somewhat in disagreement with the experimental 
result of the Trojan-horse method \cite{STP01,MPG01,TSD03}, which is 
believed to provide the bare cross sections without the influence of 
bound electrons. Thus, the problem has not been resolved completely yet. 

Besides the electron screening effects, several small effects on 
astrophysical fusion reactions have also been examined. 
These include vacuum polarization \cite{BBH97}, relativity \cite{BBH97}, 
bremsstrahlung outside the barrier \cite{BBH97}, 
atomic polarization \cite{BBH97}, radiation correction 
during the tunneling \cite{HB02}, 
zero point fluctuation of nuclei in the atom and the molecule \cite{FRVR03}, 
and the effect of finite beam width \cite{FRVR03}. 
All of these effects have been found much smaller than the screening effect. 

In this paper, we consider more corrections to astrophysical fusion reaction. 
An important fact is that the classical 
turning point of interest is much larger than the nuclear size (for instance, 
it is 288 fm for the $d+^3$He reaction at $E_{c.m.}$=10 keV), and 
effects which are relevant to the reaction have to be associated with 
the Coulomb interaction or otherwise very long ranged. The effects 
associated with the nuclear interaction will be washed out 
by a careful choice of effective nuclear potential between 
the projectile and the target, unless the energy dependence is very 
strong \cite{HB02}. In this sense, the nuclear absorption under the 
barrier \cite{H77,H79,SWW74}, which has been discussed in connection with 
the sharp rise of nuclear $S$-factor for the $^{12}$C + $^{12}$C fusion 
reaction at low energies, for instance, is not helpful for the astrophysical
reactions. One may also think about the non-local effects of the 
internuclear potential on the tunneling phenomena \cite{GR94,CPHRG97,BBR98}. 
However, this effect does not seem significant
either, as one can cast from the result of microscopic cluster model 
calculation for the $d+^3$He reaction at low energies \cite{BL90}, where the
exchange effect has been included both in the nuclear and
in the Coulomb interactions. 

As another example of small effect on fusion, 
we here consider the effect of Coulomb break-up of colliding nuclei. 
At energies which we are interested in, the break-up channel 
is most likely kinematically forbidden. However, the tunneling probability 
is still influenced through the virtual process \cite{DLV91}, and 
it is important to estimate the size of its effect for a
complete understanding of the reaction mechanism. 
This effect is also known as dipole polarizability. 
To our knowledge, this effect has not yet been computed 
in the literature, although a few calculations have 
existed based on the continuum-discretized-coupled-channels (CDCC) 
method for transfer reactions at much higher energies, which were performed 
in aiming at extracting 
the cross section of the astrophysical radiative capture 
reactions at zero incident energy \cite{FCNT99,OYIK03}. 
We also notice that the enhancement of tunneling probability 
due to the break-up coupling has been extensively discussed in the context of 
subbarrier fusion reaction of a halo nucleus 
\cite{DV94,HVDL00,DT02,HPCD93,CCDHLR02}. 

We use a three-body model in order to estimate the effect of the virtual 
Coulomb excitation of projectile on the tunneling probability. 
Denoting the coordinate between the target and the center of mass of the 
projectile by $\vec{R}$ and the coordinate between the projectile fragments 
by $\vec{r}$, the Coulomb interaction in this system reads 
\begin{eqnarray}
V_C(\vec{R},\vec{r})&=&\frac{Z_1Z_Te^2}{|\vec{R}+m_2\vec{r}/(m_1+m_2)|}
+\frac{Z_2Z_Te^2}{|\vec{R}-m_1\vec{r}/(m_1+m_2)|}, \\ 
&\sim&\frac{Z_PZ_Te^2}{R}+\frac{4\pi}{3}\,\frac{Z_Te}{R^2}\sum_\mu
Y_{1\mu}^*(\hat{\vec{R}})\,\hat{T}^{\rm E1}_\mu,
\label{Vcoup}
\end{eqnarray}
where $\hat{T}^{\rm E1}_\mu$ is the $E1$ operator given by 
\begin{equation}
\hat{T}^{\rm E1}_\mu=e_{\rm E1}\,rY_{1\mu}(\hat{\vec{r}}).
\end{equation}
Here, $e_{\rm E1}$ is the $E1$ effective charge given by 
$(m_2Z_1-m_1Z_2)e/(m_1+m_2)$, where $m_1$ and $m_2$ are the mass of the 
projectile fragments while $Z_1$ and $Z_2$ are their charges ($Z_P=Z_1+Z_2$ 
is the total charge of the projectile). For a head-on collision, the incident 
channel is a $s$-wave bound state $\phi_0(\vec{r})$ of the 
projectile coupled to the relative angular momentum $L=0$ 
for the $\vec{R}$ coordinate. This channel couples to a 
$p$-wave state $\phi_1(\vec{r})$ of the projectile via the coupling 
interaction (\ref{Vcoup}). 
The relative angular momentum for the $\vec{R}$ coordinate has to be 
$L=1$ in the 
excited state channel so that the total angular momentum is conserved. 
For simplicity, we have neglected the spin of the projectile fragments. 
The matrix element of the coupling potential between these channels 
is given by 
\begin{equation}
F(R)=\frac{\sqrt{4\pi}}{3}\frac{Z_Te^2}{R^2}\,
\sqrt{\frac{B(E1)\uparrow}{e^2}},
\end{equation}
where $B(E1)\uparrow = |\langle \phi_0||\hat{T^{\rm E1}}||\phi_1\rangle|^2$ 
is the strength of the electric dipole transition of the projectile. 

For an exponential wave function for the bound state $\phi_0$ together 
with the 
plane wave function for the scattering state $\phi_1$, a simple and 
compact expression 
for $B(E1)\uparrow$ has been derived by Bertulani, Baur, 
and Hussein \cite{BBH91}, which is given by 
\begin{equation}
\frac{dB(E1)\uparrow}{dE_\gamma}=\frac{3\hbar^2e_{\rm E1}^2}{\pi\mu_{12}^2}
\frac{\sqrt{\epsilon}(E_\gamma-\epsilon)^{3/2}}{E_\gamma^4},
\end{equation}
where $\mu_{12}=m_1m_2/(m_1+m_2)$ is the reduced mass of the projectile system 
and $\epsilon$ is the binding energy. 
This function has a peak at $E_\gamma=8\epsilon/5$, 
and the total dipole strength is given by \cite{BBH91}
\begin{equation}
B(E1)\uparrow = \frac{3\hbar^2e_{\rm E1}^2}{16\pi\mu_{12}\epsilon}.
\label{totalE1}
\end{equation}
In this work, for simplicity, we assume that the $E1$ strength is exhausted 
by a single state at $E_\gamma=8\epsilon/5$ with the strength given by 
Eq. (\ref{totalE1}). 
With this prescription, the problem is reduced to the two dimensional 
coupled-channels calculation with the coupling matrix given by 
\begin{equation}
V_{\rm coup}(R)
=\left(\matrix{0 & F(R)\cr
F(R)&E_\gamma+\frac{2\hbar^2}{2\mu R^2}\cr}\right),
\label{coup}
\end{equation}
where $\mu$ is the reduced mass for the $\vec{R}$ motion. 

In order to estimate the coupling effect, we use the adiabatic approximation 
and derive the adiabatic potential shift by diagonalizing the coupling 
matrix (\ref{coup}) at each $R$. Taking the smaller eigenvalue, the 
potential shift is given by 
\begin{equation}
\Delta V_{ad}(R)=\frac{
\left(E_\gamma+\frac{2\hbar^2}{2\mu R^2}\right)-\sqrt{
\left(E_\gamma+\frac{2\hbar^2}{2\mu R^2}\right)^2+4F(R)^2}}{2}.
\end{equation}
Note that this potential shift coincides with the adiabatic polarization 
potential which Alder {\it et al.} derived using the second order 
perturbation theory \cite{ABH56} (see also Ref. \cite{HFFB84} 
for a derivation using the Feshbach formalism),  
in the limit of $E_\gamma \gg F(R)$ and when one 
ignores the angular momentum transfer. 
We then compute the tunneling probability using the WKB formula for low 
energy, 
\begin{equation}
P(E)=\exp\left[-2\int^{R_1}_{R_0}dR\sqrt{\frac{2\mu}{\hbar^2}
(V_0(R)+\Delta V_{ad}(R)-E)}\right],
\end{equation}
where $V_0(R)=Z_PZ_Te^2/R$ is the bare Coulomb interaction, 
and $R_0$ and $R_1$ are the inner and the outer 
turning points, respectively. Notice that the adiabatic approximation provides 
the upper limit of the potential penetrability \cite{THAB94,THA95,BT98}. 
Therefore, our results should be regarded as the upper limit of the 
virtual break-up effects in the astrophysical reactions, although the 
adiabatic approximation should work well at astrophysical energies. 

The effect of the target break-up can also be taken into account in a similar 
manner. In this case, one considers five channel states: i) 
the incident channel, 
ii) the projectile break-up channel, iii) 
the target break-up channel, iv) the mutual 
break-up channel with the relative angular momentum $L=0$, 
and v) the mutual break-up channel with $L=2$. 
Both of the channels iv) and v) are coupled to the channels 
ii) and iii) by the $E1$ operator of the target and of the projectile, 
respectively. 
The adiabatic potential $\Delta V_{ad}(R)$ is given as the lowest 
eigenvalue of the 5$\times$5 coupling matrix at each $R$. 
Here we neglect the dipole-dipole term in the interaction, 
which we assume to be much smaller than the monopole-dipole term. 

Let us now numerically estimate the effect of the virtual break-up coupling on 
astrophysical fusion reactions. We first consider the effect of 
deuteron break-up on the $d+^3$He reaction. The break-up $Q$-value is 
$\epsilon$=2.22 MeV, and Eq.(\ref{totalE1}) leads to the total 
$B(E1)\uparrow$ strength of 0.558 ($e^2$ fm$^2$). The dashed line in Fig. 1 
shows the enhancement factor $f$ of the penetrability, $P(E)/P_0(E)$, 
as a function of the center of mass energy $E_{c.m.}$, where $P_0(E)$ is 
the penetrability of the bare Coulomb interaction $V_0(R)$. 
We take $R_0$ = 4.3 fm for the inner turning point \cite{BBH97}. 
We see that the enhancement factor slowly decreases as the energy decreases. 
The value of the enhancement factor is about 0.21 \% at $E_{c.m.}$ = 5.8 keV, 
which is smaller than the effect of vacuum polarization 
\cite{BBH97} by one order. 
Therefore, the effect of the dipole polarizability of deuteron 
seems to be negligible as compared to the vacuum polarization effect. 
The dotted line shows the effect of $^3$He break-up, where the binding 
energy is 5.49 MeV from the threshold of the $d+p$ system. Although this 
effect is much smaller than the deuteron break-up effect, the combined effect 
of the mutual excitations increases the penetrability in a non-negligible 
way (see the solid line). Figure 2 shows the effect of dipole break-up 
of $^7$Li nucleus (into $\alpha+t$) on the $p+^7$Li reaction. For this 
system, the $E1$ effective charge is small, and the effect of break-up is 
even 
much smaller than the $d+^3$He system. Notice that the $E1$ effective charge 
vanishes for a similar projectile, $^6$Li, which predominantly breaks into 
$\alpha +d$. 

In summary, we have studied the effect of virtual Coulomb 
break-up process of
colliding nuclei (i.e., the dipole polarizability) 
on astrophysical fusion reactions. For the deuteron
break-up, we have found that the enhancement of the tunneling
probability is about 
0.2 \% for the $d$ + $^3$He system. The effect is much smaller 
for the $^7$Li break-up in the $p$+$^7$Li system, where the
enhancement factor was found to be about 2.7$\times 10^{-3}$ \%. 
Therefore, the break-up 
effect alone does not resolve the large screening puzzle. We have a 
feeling that we have almost exhausted the list of small effects 
in astrophysical reactions. Of course, 
there are still some exotic 
effects such as the deformation of proton \cite{M03} or 
the color van der Waals force \cite{HLPB90}, but these 
effects should be extremely small in the astrophysical reaction. 
We may now be at a stage where the atomic physics based model 
has to be re-examined with a more careful and consistent treatment 
of few-body dynamics of charged particles including electrons. 

\medskip

We thank the IPN Orsay for their warm hospitality where this work was 
carried out.  
K.H. also acknowledges support from the Kyoto University Foundation. 
This work was supported in part by the U.S. National Science
Foundation Grants No. PHY-0244384 and INT-0070889 at the University of
Wisconsin.

\newpage

\begin{figure}
  \begin{center}
    \leavevmode
    \parbox{0.9\textwidth}
           {\psfig{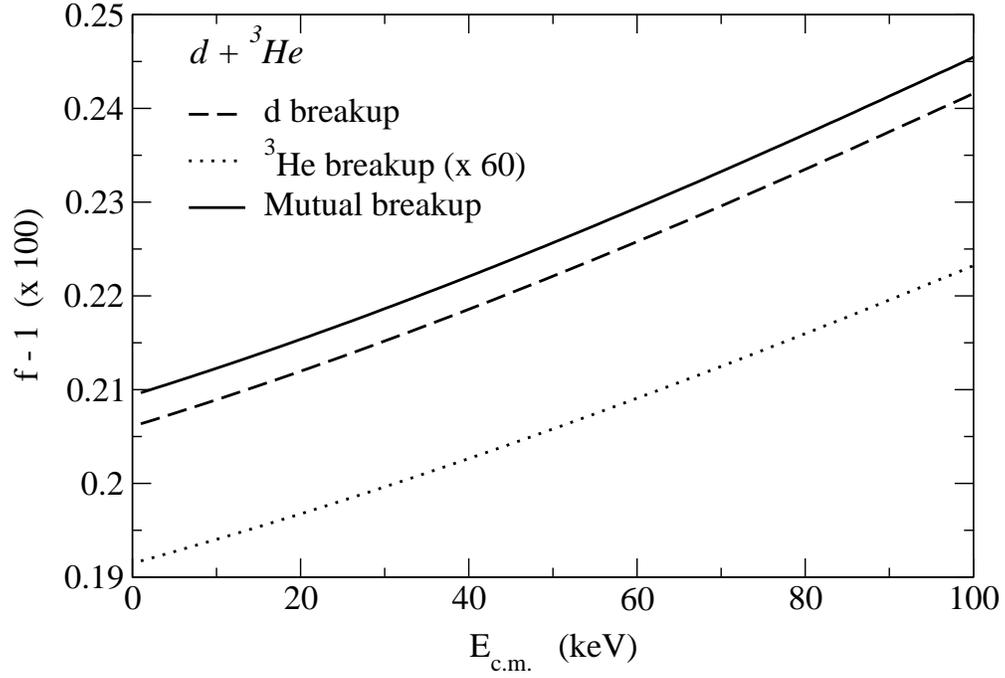}}
  \end{center}
\protect\caption{
The effect of virtual Coulomb break-up of colliding nuclei 
on the $d+^3$He reaction. 
$f$ is the enhancement factor of penetrability due to the break-up, 
measured from unity. 
The dashed and the dotted lines show the effect of break-up of the $d$ and 
the $^3$He nuclei, respectively. The solid line is the combined effect 
of mutual break-up of both the projectile and the target nuclei. }
\end{figure}

\newpage

\begin{figure}
  \begin{center}
    \leavevmode
    \parbox{0.9\textwidth}
           {\psfig{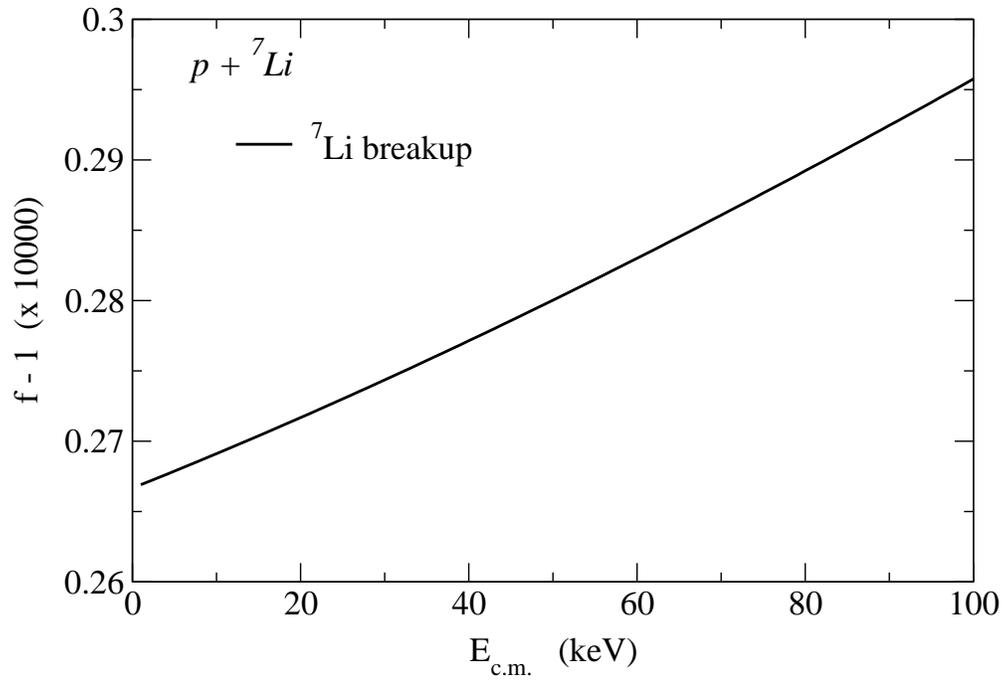}}
  \end{center}
\protect\caption{
Same as fig. 1, but for the effect of $^7$Li break-up 
on the $p+^7$Li reaction. 
}
\end{figure}

\end{document}